\documentclass[a4paper,8pt]{article}
\usepackage[latin1]{inputenc}

\newtheorem{thm}{Théorème}[section]

\newtheorem{rem}[thm]{Remark}

\newtheorem{rul}[thm]{Rule}
\newcommand{\psla}{\mbox{$\not{\! p}$}}

\newcommand{\caln}{{\cal N}}

\def\dem {\noindent {\bf Proof : }}

\def\sqw{\hbox{\rlap{\leavevmode\raise.3ex\hbox{$\sqcap$}}$\sqcup$}}
\def\findem{\ifmmode\sqw\else{\ifhmode\unskip\fi\nobreak\hfil
\penalty50\hskip1em\null\nobreak\hfil\sqw
\parfillskip=0pt\finalhyphendemerits=0\endgraf}\fi}

\usepackage{latexsym}
\usepackage{amsmath,amssymb,amsfonts,graphicx}

\newcommand{\dsp}{\displaystyle}

\begin{document}

\begin{titlepage}

\begin{flushright}
LAPTH-1218\\
November 2007
\end{flushright}
\vspace{1.cm}

\begin{center}
{\LARGE\bf Six-Photon Amplitudes in Scalar QED}\\[1cm]%

{C.~Bernicot$^{a}$, J.-Ph.~Guillet$^{a}$}\\[1cm]%

{\em $^{a}$LAPTH, Universit\'e de Savoie, CNRS\\
 B.P. 110, F-74941 Annecy-le-Vieux Cedex, France.}\\[.5cm]

\end{center}
\normalsize

\vspace{1cm}

\begin{abstract}

The analytical result for the six-photon helicity amplitudes in
scalar QED is presented. To compute the loop, a recently developed
method based on multiple cuts is used. The amplitudes for QED and
$QED^{\caln=1}$ are also derived using the supersymmetric
decomposition linking the three theories.
\end{abstract}

\vspace{5cm}

version \today

\end{titlepage}

\section{Introduction and Notation}

\hfil

The light-by-light scattering is a prediction of quantum
electrodynamic despite the fact that it has never been observed so
far. The four-photon amplitudes, in QED, have been computed in the
fifties at one loop for massive fermion \cite{karplus} and
recently at two loops for massless fermions in QED \cite{gamtwo}
and in $\caln=1$ supersymmetric QED \cite{gamsusy}. The first
result for the six-photon amplitudes, at one loop in QED and for
massless fermions, was obtained for the MHV (Maximal Helicity
Violating) amplitude by Mahlon \cite{mahlon2}. The complete
helicity amplitudes, in QED, has been computed numerically by
direct integration of the Feynman diagrams \cite{nagy}, and also,
by using reduction at the integrand level
\cite{papadopoulos:6photon}. In the same time, it has also been
computed analytically using unitary cut methods and cross checked
with a reduction method \cite{binoth:2007}. A compact formula has
been given, proving the power of the unitary cut methods. The
unitarity-cut methods were developed first in \cite{Bern:1994zx}
and then in \cite{Britto:2004nc} and are currently under intense
developments
\cite{Britto:2005ha,Britto:2006sj,Mastrolia:2006ki,method7,method7bis}.
\hfil

\hfil

Although the six-photon amplitudes are out of reach for nowadays
experiments, they provide a good laboratory reaction to settle
efficient methods to compute one loop multi-leg amplitudes.
Indeed, multi-particle processes involving Quantum ChromoDynamics
(QCD) will play an important role in the physics probed by the
hadronic colliders at the TeV scale. In particular at the future
Large Hadron Collider (LHC), the production  of four, and even
five or six jets will not be marginal. Besides providing a refined
probe of the dynamics of colour, such QCD processes constitute a
background to the search for new particles. Indeed, the search for
many of these new particles at hadronic colliders often relies on
signatures based on cascade decays. The latter end up with final
states involving a large jet multiplicity. Furthermore, the lowest
order estimates for such processes are plagued by the well-known
deficiencies of large renormalization and factorization scale
dependencies, poor multi-jet modelling and large sensitivity to
kinematic cuts. Therefore the calculation of next-to-leading-order
(NLO) corrections to such cross sections is a necessary step
forward.

\hfil

In this article, for scalar QED (respectively QED, $\caln=1$ supersymmetric QED) a generic $N$-photon helicity amplitude is denoted by: $A_{N}^{scalar}$ (respectively $A_{N}^{fermion}$, $A_{N}^{\caln=1}$). We will use unitarity-cut methods to compute the six-photon amplitudes in scalar QED. From these results, we can derive results for QED and $\caln=1$ supersymmetric QED. To achieve this, we use a relation which relates the three theories. To find it, we proceed as follows. Starting with the $N$-photon QED amplitudes and using the fact that degrees of freedom for internal lines can be added and subtracted \cite{Dixon:TASI}, we can write the following relation:
\begin{align}
    A_{N}^{fermion} = -2 A_{N}^{scalar} + A_{N}^{\caln=1} \label{supersymetricdecomposition}
\end{align}

\hfil

\noindent To calculate those amplitudes, we use the spinor
helicity formalism developed in ~\cite{spinor:chinois}. For the
spinorial product, we introduce the following notation:
\begin{align}
    \langle p_{a}- | p_{b}+ \rangle & = \langle ab \rangle &\\
    \langle p_{a}+ | p_{b}-\rangle &= [ab] &\\
    \langle p_{a}- | \psla_{b} | p_{c} - \rangle &= \langle abc
    \rangle = [cba] = \langle p_{c}+ | \psla_{b} | p_{a} + \rangle &\\
    \langle p_{a}+ | \psla_{b} \, \psla_{c} | p_{d} - \rangle & = [ abcd ] = - [ dcba ] = -\langle p_{d}+ | \psla_{c} \, \psla_{b} | p_{a} - \rangle &
\end{align}
\noindent Moreover we use $p_{i...j} = p_{i}+ ... + p_{j}$ and
$s_{i..j} = (p_{i}+ ... + p_{j})^{2} = p_{i...j}^{2}$.

\hfil

\noindent The outline of the paper is as follows. In section 2, we
analyse the structure of the amplitudes, and compute the different
tree amplitudes necessary for our calculation. In section 3, we
give an analytical result for the six-photon amplitudes in scalar
QED and in section 4, we derive analytical results for QED and
$\caln=1$ supersymmetric QED. In section 5, we plot the different
amplitudes for some kinematics and discuss potential problems.

\hfil

\section{Structure of the amplitudes}

\hfil

\subsection{Decomposition of the amplitudes $A_{6}^{fermion}$, $A_{6}^{scalar}$ and $A_{6}^{\caln=1}$}

\hfil

The standard reduction methods, for example \cite{van Neerven:1983vr,Bern:1993kr,method1} show that
any amplitude can be written as a combination of master
integrals. This set of master integrals is not unique.

\hfil

From now, we consider theory with massless particles. In the case
of the six-photon amplitudes, we use the following decomposition:
\begin{align}
    A_{6} =  \dsp \sum_{i \in \sigma(1,2,3,4,5,6)} & \left(  a_{i} \, F_{4} + b_{i} \, F_{3} + d_{i} \, F_{2A} + c_{i} \, F_{2B} + e_{i} \, F_{1}+ f_{i} \, F_{0} \right. & \nonumber \\
    &  \left. \dsp + g_{i} \, {I_{3}^{n}}^{3mass} + h_{i} \, {I_{3}^{n}}^{2mass} + i_{i} \, {I_{3}^{n}}^{1mass}  +  j_{i}\, I_{2}^{n} + \textrm{rational
    terms}\right)
    & \label{decompositiongood}
\end{align}
\noindent where $F_{4}$ (respectively $F_{3}$, $F_{1}$ and
$F_{0}$) is the so called "finite part" of the four point
function, in $n$ dimensions, with 4 external masses (respectively
three external masses, one external mass and zero external mass),
$F_{2A}$ (respectively $F_{2B}$) the so called "finite" part of
the $n$ dimensional four point function with two adjacent external
masses (respectively with two opposite external masses). Only this
set of functions $\left\{ F_{2A},F_{2B}, F_{1} \right\}$ will be
used, their exact definition can be found, for example in
\cite{Binoth:2001vm}, and to be self consistent we recall them in
the appendix \ref{scalar_int}. In addition, ${I_{3}^{n}}^{3mass}$ (respectively
${I_{3}^{n}}^{2mass}$, ${I_{3}^{n}}^{1mass}$) is the $n$ dimensional
three point function with three external masses (respectively two
external masses, one external mass) and $I_{2}^{n}$ is the $n$
dimensional two point function. The IR divergences are carried by
the function ${I_{3}^{n}}^{2m}$ and ${I_{3}^{n}}^{1m}$ and the UV
one by the function $I_{2}^{n}$.

\hfil

\noindent Using unitary-cut methods, we only have to compute the
coefficients $ a_{i}... j_{i}$ and rational terms. Most of them
are related by Bose symmetry or parity.

\hfil

\subsection{Tree amplitudes}

\hfil

In the framework of unitary-cut methods, we need to compute first
tree amplitudes. In this subsection, we will present only the tree
amplitudes in QED and scalar QED useful for our six-photon
amplitude computation. The needed tree amplitudes, are the
amplitudes corresponding to the reactions: two scalars (fermions)
into $N$-photons with the same helicity and two scalars (fermions)
into $N$-photons with the same helicity but one. We assume that
all the photons are ingoing, and $p_a$ and $p_b$ are the four
momentum of the scalars (fermions):
\begin{figure}[httb!]
\centering
\includegraphics[width=9cm]{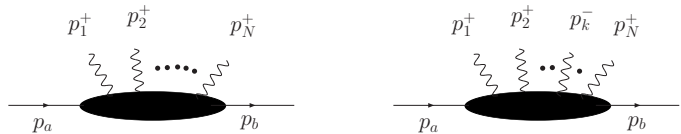}\\
\caption{\scriptsize \textit{ Tree amplitudes needed for the six-photon amplitudes. The
particles associated with a plain line are scalars or fermions.}}
\end{figure}

\begin{align}
    A_{tree}^{scalar}( 1^{+},..., N^{+}) & =0 & \label{arbre1QEDS} \\
    A_{tree}^{fermion}( 1^{+},..., N^{+}) & =0 & \label{arbre2QEDS} \\
    A_{tree}^{scalar}( 1^{+},..., N^{+},k^{-}) & = \dsp i \left(e \sqrt{2} \right)^{N+1} \sum _{\sigma(\{1..N\} \backslash k)} \frac{1}
    { \langle 12 \rangle \langle 23 \rangle ... \langle N-1 N\rangle} \frac{\langle ka1\rangle}{\langle 1a1\rangle} \frac{\langle kbN \rangle}{\langle NbN \rangle} & \nonumber \\
    & = \dsp i \left(e \sqrt{2} \right)^{N+1} \frac{\langle ka \rangle \langle kb \rangle}{\prod _{j=1, j \ne k}^{N} \langle ja \rangle \langle jb
    \rangle}\langle ab \rangle ^{N-1}  = ie \sqrt{2}\frac{\langle ka \rangle \langle kb \rangle}{\langle ab \rangle} \prod_{i=1, i \ne k}^{N} S_{i}& \label{arbre3QEDS} \\
    A_{tree}^{fermion}( 1^{+},..., N^{+},k^{-}) & = A_{tree}^{scalar}( 1^{+},..., N^{+},k^{-})
    \left(\frac{\langle ka \rangle}{\langle kb \rangle} + \frac{\langle kb \rangle}{\langle ka \rangle}\right) \label{arbre4QEDS} &
\end{align}
\noindent where $\dsp S_{i} = -e \sqrt{2}\frac{\langle ab
\rangle}{\langle ai \rangle \langle ib \rangle}$ is the eikonal
factor. Note that in equation (\ref{arbre4QEDS}), a sum over the helicities of the fermions has been performed.

\hfil

\subsection{Additional reductions} \label{structure}

\hfil

Now using the properties of the QED theories, we can simplify
furthermore the decomposition $\left( \ref{decompositiongood}
\right)$ of the six-photon amplitudes. Those comments and rules is
available only for these amplitudes.

\begin{rem} \label{IRdiverg}
Only the functions ${I_{3}^{n}}^{2m},{I_{3}^{n}}^{1m}$ are IR divergent. Since each diagram is not
IR divergent, the coefficients $h_{i}$ and $i_{i}$ are zero.
\end{rem}

Each Feynman diagram of the six-photon amplitudes is free of IR divergences thanks to the numerator
of the fermionic propagator in QED or the structure of the vertex
in scalar QED. If the reduction is done by pinching propagators,
we get sub-diagrams which are not IR divergent. After the
reduction, we obtain three point sub-diagrams and the "finite"
part of four point scalar integrals. Since the three point
sub-diagrams are free of IR divergences, they cannot be expressed
in term of one mass/two mass three point scalar integrals
${I_{3}^{n}}^{2m}$, ${I_{3}^{n}}^{1m}$ and so the coefficients $h_{i}$ and $i_{i}$
are zero \cite{method1}.

\begin{rem} \label{UVandrational}
Using standard reduction (for example \cite{method1}), we can show
for QED and scalar QED, that the coefficients in front of two
point functions are zero and also that the rational terms are zero.
\end{rem}

The first statement is in accordance with the fact that each
diagram of the six-photon amplitudes is free of UV divergences.
From that, it is not obvious that the different coefficients are
zero, we have proven it by explicit calculation. As a consequence
of remarks (\ref{IRdiverg}) and (\ref{UVandrational}), there are
no logarithmic terms in the six-photon amplitudes. The second
statement, shown in ref. \cite{binoth2}, is also far from being
obvious. Indeed, according to power counting arguments
\cite{BDDK}, the rational terms can be present. In fact, they are
present for individual Feynman diagram but these rational terms
sum up to zero when adding
all the diagrams.\\

Now, from what has been said previously, we can reduce the
decomposition of the amplitude $\left( \ref{decompositiongood}
\right)$: for the six-photon case, the coefficients
$h_{i},i_{i}$, and $j_{i}$ are zero and the rational term is also nul.
In addition, since we have only six photons on shell, the
coefficients $a_{i}$, $b_{i}$ and $f_i$ are also zero. So each
amplitude $A_{6}^{fermion}$, $A_{6}^{scalar}$ can be written as:
\begin{align}
    A_{6} =   \dsp \sum_{i \in \sigma(1,2,3,4,5,6)} & \   d_{i} \, F_{2A} + c_{i} \, F_{2B} + e_{i} \, F_{1} + g_{i} \, {I_{3}^{n}}^{3mass} \label{combination}
\end{align}

\hfil

\noindent From the helicity structure of trees, we can derive some
rules which will reduce furthermore the decomposition
(\ref{combination}).

\setcounter{thm}{0}

\begin{rul}
Consider a master integral with mass. If the mass is formed only
with photons with the same helicity, therefore the coefficient in
front of the master integral is zero. \label{rule1}
\end{rul}

\dem Using the cut technics, we get that the coefficient in front
of this integral is proportional to the tree amplitude which, once
pinched, yields the mass. Formulae
(\ref{arbre1QEDS},\ref{arbre2QEDS}) show that the on-shell tree
amplitudes with photons having the same helicity are zero. So the
coefficient of a master integral with a mass formed by photons
with the same helicity is zero. \hfil

\begin{rul}
For the box with one mass and with two adjacent masses, the
helicity of two adjacent massless legs must be alternate. In the
case of the box with two opposite masses, the helicities of the
two opposite massless legs must be the same. If it is not the
case, the coefficient in front of the master integral is zero.
\label{rule2}
\end{rul}

\dem In the case of the one mass box and the two adjacent mass
box, if the helicities of two adjacent massless legs are the same,
the coefficient, in front of the box, will be proportional to
trees, eqs. (\ref{arbre1QEDS},\ref{arbre2QEDS}), which are zero. The case of the two opposite mass box is
more complicated and a proof is given in the appendix
\ref{p_rule2}.

\hfil

Now we can further reduce the decomposition of all helicity
amplitudes thanks to these last rules. We begin with the most
simple amplitudes: $A_{N}(1^{\pm},2^{+}...,N^{+})$. With at most
one negative photon, we can have only one mass according to the
rule (\ref{rule1}), so $c_{i},d_{i},g_{i}=0$. But in this case,
the helicities cannot be alternate so, from the rule
(\ref{rule2}), we deduce that each $e_{i}=0$. Therefore, we get
for these amplitudes:
\begin{align}
    \forall N> 4, & \ \  A_{N}^{scalar}(1^{\pm},2^{+}...,N^{+}) =0 &\\
    \forall N> 4, & \ \  A_{N}^{fermion}(1^{\pm},2^{+}...,N^{+}) =0 & \\
    \forall N> 4, & \ \  A_{N}^{\caln=1}(1^{\pm},2^{+}...,N^{+}) =0 &
\end{align}
\noindent These results were already found by Mahlon \cite{mahlon2} many
years ago.

\hfil

Then we study the MHV amplitude $
A_{6}(1^{-},2^{-},3^{+},4^{+},5^{+},6^{+})$. We have only two
negative photons so we can only have at most two masses:
$g_{i}=0$. The helicities must be alternate according to rule
(\ref{rule2}) therefore for the MHV amplitude, $d_{i}=0$. So we
have:
\begin{align}
    A_{6}(--++++) = i\frac{\left( e\sqrt{2}\right) ^{6}}{16 \pi^{2}} \dsp \sum_{\sigma(1,2)} \sum_{\sigma(3,4,5,6)} & \  \frac{c_{i}}{2} F_{2B}(s_{215},s_{415},s_{15},s_{26}) + \frac{e_{i}}{2} F_{1}(s_{23},s_{24},s_{156})  \label{combination--++++}
\end{align}
\noindent Some permutations leave master integrals invariants, that
is why we divide the coefficient by the adequate number.

\hfil

\noindent It is rather easy to show that, in the case of the MHV
six-photon amplitude, the two coefficients $d_{i}$ and $e_{i}$ are
related. Indeed, we can take the limit such that the photon with
helicity "+" forming one of the mass of the two opposite mass box
is soft. In this limit, we have the following relation: $ \lim
_{m_{2} \rightarrow 0} F_{2B} = F_{1}$. Since the MHV amplitude is
composed by two MHV trees already expressed in term of eikonal
factors eq. (\ref{arbre3QEDS}) and since the five photon amplitudes are
zero, we can deduce that $ c_{i} = - e_{i}$. Finally, we have the
following decomposition for the MHV amplitude:
\begin{align}
    A_{6}(--++++) = \dsp i\frac{\left( e\sqrt{2}\right) ^{6}}{16 \pi^{2}} \sum_{\sigma(1,2)} \sum_{\sigma(3,4,5,6)} & \
    \frac{c_{i}}{2} \left( F_{2B}(s_{315},s_{415},s_{15},s_{26}) - F_{1}(s_{23},s_{24},s_{156}) \right) \label{combination--++++}
\end{align}
\noindent We have to compute only one coefficient.

\hfil

Lastly, we examine the form of the decomposition of the Next to
MHV amplitude (NMHV) $
A_{6}(1^{-},2^{-},3^{-},4^{+},5^{+},6^{+})$. Here we can have
three mass three point functions. But the rule (\ref{rule2})
imposes $c_{i} = 0$. So we have only three coefficients to
calculate and the amplitude can be written as:
\begin{align}
    A_{6}(---+++) = \dsp i\frac{\left( e\sqrt{2}\right) ^{6}}{16 \pi^{2}} \sum_{\sigma(1,2,3)} \sum_{\sigma(4,5,6)}& \ 2 \ d_{i}
    F_{2A}(s_{14},s_{452},{s}_{25},{s}_{36})
    + \frac{e_{i}}{2} F_{1}(s_{63},s_{61},s_{425})& \nonumber \\
    & + \dsp \frac{{e_{i}}^{*}}{2} F_{1}(s_{25},s_{24},s_{136})
    + \frac{g_{i}}{6}{I_{3}^{n}}^{3m}({s}_{14},{s}_{25},{s}_{36})
 \label{combination---+++}
\end{align}
\noindent Here again, we divide the coefficients by the adequate
number to take into account permutations leaving invariant the
corresponding master integrals. There are two kinds of one mass
box integrals. The first kind has two photons with a positive
helicity and one with a negative helicity forming the mass,
whereas the second kind has one photon with a positive helicity and
two photons a with negative helicity forming the mass. As we have
three photons with a negative helicity and three with a positive
helicity, the two kinds of one mass box integrals are directly
related by parity. That is why the coefficient in front of them
are complex conjugate.

\hfil

\noindent In the next section, we will give the results for the various cases.

\hfil

\section{$A_{6}^{scalar}$ amplitudes}

\hfil

\subsection{$ A_{N}^{scalar}(1^{-},2^{-},3^{+}...,N^{+}), \ N> 4$ helicity amplitudes}

\hfil

In this section, we calculate the MHV amplitude
$A_{N}^{scalar}(1^{-},2^{-},3^{+}...,N^{+})$ for $N$ photons with
$N>4$. The generalization to $N$ photons is rather easy, but note
that we have to consider $N>4$ otherwise there are some UV
problems. Using the quadruple cut technics \cite{Britto:2004nc},
by direct computation, we obtain:
\begin{align}
     A_{N}^{scalar} & (--+...+) = \dsp \ \ i\frac{\left( e\sqrt{2}\right) ^{N}}{16 \pi^{2}}\sum_{\sigma(1,2)} \sum_{\sigma(3..N)}
    \frac{d_i^{scalar}}{(N-4)!} F_{1}(s_{23},s_{24},s_{15...N}) & \nonumber \\
    & + \dsp i\frac{\left( e\sqrt{2}\right) ^{N}}{16 \pi^{2}} \sum_{\sigma(1,2)} \sum_{\sigma(3..N)}\sum_{M=5}^{N-1} \frac{(-1)^{M-6}d_i^{scalar}}{(N-M)!(M-4)!}
    F_{2B}(s_{135...M},s_{145...M},s_{15...M},s_{2M+1...N}), &
\end{align}
\noindent where
\begin{equation}
d_i^{scalar} = - \dsp \frac{\langle 34
\rangle^{N-6}}{\prod_{j=5}^{N}\langle 3j\rangle \langle 4j\rangle}
\frac{\langle 13 \rangle \langle 41 \rangle \langle 23\rangle
\langle 42 \rangle}{s_{34}} \frac{[34]}{\langle 34 \rangle }
\end{equation}
The factorial coefficient $(N-M)!(M-4)!$ and $(N-4)!$ are the
number of permutations which leaves invariant the mass of the master
integral. The formula obtained is explicitly invariant by exchange
of the two photons with a negative helicity. In the case where
$N=6$, the amplitude reduces to:
\begin{align}
    A_6^{scalar}(--++++) = \dsp - i\frac{\left( e\sqrt{2}\right) ^{6}}{16 \pi^{2}} \sum_{\sigma(1,2)} \sum_{\sigma(3..6)}
    \dsp
    \frac{\langle 13 \rangle \langle 41 \rangle \langle 23\rangle \langle 42 \rangle}{\langle 35\rangle
    \langle 45\rangle \langle 36\rangle \langle 46\rangle } \frac{[34]}{\langle 34 \rangle }
    \frac{F_{1}(s_{23},s_{24},s_{156})
    -F_{2B}(s_{135},s_{145},s_{15},s_{26})}{s_{34}} \label{mhv_scalar}&
\end{align}
\noindent It remains only one Gram determinant $\left( s_{34}
\right)$ in the denominator but $F_{1}$, $F_{2B}$ $\simeq s_{34}$
when $s_{34} \rightarrow 0$. Therefore the potential numerical
problem when the Gram determinant vanishes is under control.

\hfil

\subsection{$ A_{6}^{scalar}(1^{-},2^{-},3^{-},4^{+},5^{+},6^{+})$ helicity amplitude}

\hfil

We use the quadruple-cut technics to calculate the box
coefficients~\cite{Britto:2004nc}, and the triple-cut technique to
calculate the triangle coefficients~\cite{Forde:2007mi}. We obtain
for the amplitude:
\begin{align}
    A_{6}^{scalar}(---+++) = \dsp i\frac{\left( e\sqrt{2}\right) ^{6}}{16 \pi^{2}} \sum_{\sigma(1,2,3)}\sum_{\sigma(4,5,6)} & 2 \ d_{i}^{scalar}
    F_{2A}(s_{14},s_{452},{s}_{25},{s}_{36})
    + \frac{e_{i}^{scalar}}{2} F_{1}(s_{63},s_{61},s_{425})& \nonumber \\
    & + \dsp \frac{e_{i}^{scalar}}{2}^{*}F_{1}(s_{25},s_{24},s_{136})
    + \frac{g_{i}^{scalar}}{6} {I_{3}^{n}}^{3m}({s}_{14},{s}_{25},{s}_{36})
    &
\end{align}
\noindent where
\begin{align}
    d_{i}^{scalar} &=  \dsp -  \frac{s_{425} \langle 24 \rangle [16]}{\langle 45 \rangle [31] [1p_{425}4]^{2} }
    \frac{[1p_{425}2][6p_{425}4]}{[1p_{425}5][3p_{425}4]} &\\
    e_{i}^{scalar} &= -  \frac{\langle 2 p_{425} 1 \rangle \langle 2 p_{425} 3 \rangle  [36][16] s_{425}}
    {\langle 4p_{425}1 \rangle \langle 5p_{425}3 \rangle \langle 5p_{425}1 \rangle \langle 4p_{425}3 \rangle }\frac{\langle 31 \rangle}{ [31] s_{31}}  &\\
    {e_{i}^{scalar}}^{*} &= - \frac{[6 p_{613}5][6p_{613}4] \langle
    42\rangle\langle 52\rangle s_{613}}{[1p_{613}4][1p_{613}5][3p_{613}4][3p_{613}5]}\frac{[45]}
    {\langle 45 \rangle s_{54}} &\\
    g_{i}^{scalar} & = \dsp  \frac{[4 p_{25} 1]}{[1 p_{25} 4]} \frac{[5 p_{14} 2]}{[2 p_{14} 5]} \frac{[6 p_{25}3]}{[3 p_{25}6]} \sum_{\gamma_{\pm}}
    \frac{[1K_{2}^{b}1]}{[4K_{2}^{b}4]}\frac{[2K_{2}^{b}2]}{[5K_{2}^{b}5]}\frac{[3K_{2}^{b}3]}{[6K_{2}^{b}6]}
    &\\
    {K_{2}^{b}}^{\mu} & = \gamma_{\pm} \left(-p_{25}\right)^{\mu} - s_{25} \left( p_{14} \right) ^{\mu}& \\
    \gamma_{\pm} & = -p_{25}.p_{14} \pm \sqrt{\Delta} & \\
    \Delta & = (p_{25}.p_{14})^{2} - p_{14}^{2}p_{25}^{2} &
\end{align}
\noindent The result is very compact. By expanding the coefficient
$g_{i}^{scalar}$, we find that all square roots, coming from
$\gamma_{\pm}$, disappear as it should be. All the coefficients
are rational functions of spinor products of external momenta.
There is also a Gram determinant in the denominator of the
coefficients $d_i^{scalar}$ and $g_i^{scalar}$. These two Gram
determinants go to zero in the same phase space region. For this region, in the numerator, it is a combination of $F_{2A}$ and
${I_3^n}^{3m}$ which cancels in such way that there is no
singularity (for more details see \cite{method1}). We do not give
here the explicit formulae because they break the simplicity of
the expressions but we implement them in our numerical code. It is
particulary important to do this for the scanning of the Landau
singularities.

\hfil

In the next section, we present the result for the amplitudes with
a fermion loop and a sfermion loop.

\hfil

\section{$A_{6}^{fermion}$ and  $A_{6}^{\caln=1}$ amplitudes}

\hfil

\subsection{$ A_{N}^{fermion}(1^{-},2^{-},3^{+}...,N^{+})$ and $ A_{N}^{\caln=1}(1^{-},2^{-},3^{+}...,N^{+})$ helicity amplitudes}

\hfil

To calculate the MHV amplitudes : $A_{N}^{fermion}(1^{-},2^{-},3^{+}...,N^{+})$ and $A_{N}^{\caln=1}(1^{-},2^{-},3^{+}...,N^{+})$, we use extensively the formula $ \left( \ref{arbre4QEDS} \right)$ at the integrand level. The idea is the following. The introduction of $ \left( \ref{arbre4QEDS} \right)$ under the integral, in the cut fermion amplitude ${A^{fermion}_{N}}_{cut}$, allows to write the cut fermion amplitude as the cut scalar amplitude ${A_{N}^{scalar}}_{cut}$ plus some terms. Using the supersymmetric decomposition $\left( \ref{supersymetricdecomposition} \right)$ the remaining part is identified with the cut supersymmetric amplitude ${A_{N}^{\caln=1}}_{cut}$. So, we do not have to calculate all the supersymmetric diagrams.

\hfil

\begin{figure}[httb!]
\centering
\includegraphics[width=6cm]{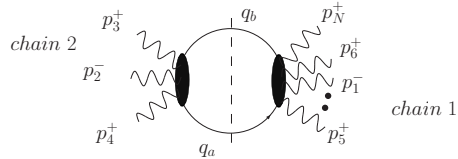}
\caption{\scriptsize \textit{Spinor QED loop.}} \label{Figure2}
\end{figure}

\noindent We show how it works on an example. We consider a
fermion loop with two cuts on propagators $q_{a}$ and $q_{b}$. The
figure \ref{Figure2} shows how the helicities are shared: two
trees with one photon with a negative helicity.
We compute the cut amplitude ${A^{fermion}_{N}}_{cut}$ corresponding to figure
\ref{Figure2}:
\begin{align}
    {A^{fermion}_{N}}_{cut} & = - \dsp \int d^{n}q \, \delta \left( q_a^{2} \right) \delta \left( q_b^{2} \right)
    \langle b \left( \Pi_{+} + \Pi_{-} \right) \textrm{(chain1)} a\rangle
    \langle a \left( \Pi_{+} + \Pi_{-} \right) \textrm{(chain2)}
    b\rangle & \nonumber \\
    & = - \dsp \int d^{n}q \, \delta \left( q_a^{2} \right) \delta \left( q_b^{2} \right)
    \left( \frac{\langle 1b \rangle \langle 2a \rangle}{\langle 1a \rangle \langle 2b \rangle} + \frac{\langle 1a \rangle \langle 2b \rangle}{\langle 1b \rangle \langle 2a \rangle} \right)
    {A^{scalar}_{1}}_{tree}{A^{scalar}_{2}}_{tree}  & \nonumber \\
    & = -2 \int d^{n}q \, \delta \left( q_a^{2} \right) \delta \left( q_b^{2} \right){A^{scalar}_{1}}_{tree}{A^{scalar}_{2}}_{tree} -\langle 12 \rangle ^{2} \int d^{n}q \, \delta \left( q_a^{2} \right) \delta \left( q_b^{2} \right) \frac{s_{ab}^{2}{A^{scalar}_{1}}_{tree}{A^{scalar}_{2}}_{tree}}{\langle 1ab1 \rangle \langle
    2ab2
    \rangle} \label{etape} &
\end{align}
where $ \Pi_{\pm} = \frac{1 \pm \gamma_{5}}{2}$ are the chiral
projectors. We recognize the cut scalar amplitude ${A_{N}^{scalar}}_{cut}$ in
the first term of the right hand side of equation (\ref{etape}).
Since the cut amplitude ${A^{fermion}_{N}}_{cut}$ also obeys to the
supersymmetric decomposition $\left(
\ref{supersymetricdecomposition} \right)$, that means that the
second term of the left hand side of eq. $\left( \ref{etape}
\right)$ is identified as the cut supersymmetric amplitude. Using this
trick, we can easily calculate the supersymmetric amplitude and
obtain straightforwardly the spinor amplitude. For these two
amplitudes, we get:
\begin{align}
    A_{N}^{fermion/\caln=1} & (--+...+) =  \ \ \dsp i\frac{\left( e\sqrt{2}\right) ^{N}}{16 \pi^{2}}\sum_{\sigma(1,2)} \sum_{\sigma(3..N)}
    \frac{d_i^{fermion/\caln=1}}{(N-4)!} F_{1}(s_{23},s_{24},s_{15...N})
   & \nonumber \\
   & \dsp + i\frac{\left( e\sqrt{2}\right) ^{N}}{16 \pi^{2}} \sum_{\sigma(1,2)} \sum_{\sigma(3..N)} \sum_{M=5}^{N-1} \frac{(-1)^{M-6}d_i^{fermion/\caln=1}}{(N-M)!(M-4)!}
    F_{2B}(s_{135...M},s_{145...M},s_{5...M},s_{M+1...N}), &
\end{align}
\noindent where $\dsp d_i^{fermion} = 2\frac{\langle 34
\rangle^{N-6}}{\prod_{j=5}^{N}\langle 3j\rangle \langle 4j\rangle}
\frac{\langle 13\rangle ^{2} \langle 42 \rangle ^2}{s_{34}}
\frac{[34]}{\langle 34 \rangle } $, and $\dsp d_i^{\caln=1} =
-\frac{\langle 34 \rangle^{N-6}}{\prod_{j=5}^{N}\langle 3j\rangle
\langle 4j\rangle} \langle 12\rangle ^2 $. In the case of six
photons, we get:
\begin{align}
    A_{6}^{fermion}(--++++) = \dsp i\frac{\left( e\sqrt{2}\right) ^{6}}{16 \pi^{2}} \sum_{\sigma(1,2)} \sum_{\sigma(3..6)}
    \dsp
    \frac{2\langle 13\rangle ^{2} \langle 42 \rangle ^2}{\langle 35\rangle
    \langle 45\rangle \langle 36\rangle \langle 46\rangle }\frac{[34]}{\langle 34 \rangle }
    \frac{F_{1}(s_{23},s_{24},s_{156}) -F_{2B}(s_{135},s_{145},s_{15},s_{26})
    }{s_{34}} \label{mhv_fermion}& \\
    A_{6}^{\caln=1}(--++++) = \dsp i\frac{\left( e\sqrt{2}\right) ^{6}}{16 \pi^{2}} \sum_{\sigma(1,2)} \sum_{\sigma(3..6)}
    \dsp
    \frac{-\langle 12 \rangle ^{2}}{\langle 35\rangle
    \langle 45\rangle \langle 36\rangle \langle 46\rangle}
    \left( F_{1}(s_{23},s_{24},s_{156}) -F_{2B}(s_{135},s_{145},s_{15},s_{26})
    \right)  \label{mhv_n=1}&
\end{align}
\noindent Full agreement is found with \cite{mahlon2,binoth:2007}
for the fermion amplitude $A_6^{fermion}(--++++)$. We can do the same
comments as in scalar QED. We point out that the Gram determinants
$s_{34}$ have disappeared in the supersymmetric amplitude
$A_{6}^{\caln=1}(--++++)$.

\hfil

\subsection{$ A_{6}^{fermion}(1^{-},2^{-},3^{-},4^{+},5^{+},6^{+})$ and $ A_{6}^{\caln=1}(1^{-},2^{-},3^{-},4^{+},5^{+},6^{+})$ helicity amplitudes}

\hfil

To calculate those two amplitudes, we use again quadruple cut
technics~\cite{Britto:2004nc} and triple cut
technics~\cite{Forde:2007mi}. As in the preceding subsection, we
use the supersymmetric decomposition $\left(
\ref{supersymetricdecomposition} \right)$ to extract the
supersymmetric amplitude and then obtain the fermion one.

\begin{figure}[httb!]
\centering
\includegraphics[width=6cm]{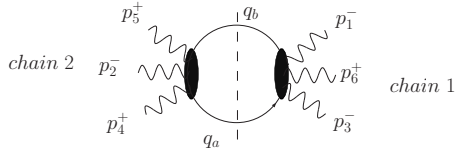}
\caption{\scriptsize \textit{Spinor QED loop.}} \label{Figure3}
\end{figure}

\noindent As in the previous subsection, we treat an example to
illustrate how it works. It is a bit different here because the
trees do not have the same helicity structure. We consider a
fermion loop with two propagators cut as shown on figure
\ref{Figure3}. If we compute the corresponding cut amplitude, we
get:
\begin{align}
    {A^{fermion}_{N}}_{cut} & = - \dsp \int d^{n}q \, \delta \left( q_a^{2} \right) \delta \left( q_b^{2} \right)
    \langle b \left( \Pi_{+} + \Pi_{-} \right) \textrm{(chain1)} a\rangle
    \langle a \left( \Pi_{+} + \Pi_{-} \right) \textrm{(chain2)}
    b\rangle & \nonumber \\
    & = -2 \int d^{n}q \, \delta \left( q_a^{2} \right) \delta \left( q_b^{2}
    \right){A^{scalar}_{1}}_{tree}{A^{scalar}_{2}}_{tree} -\langle
    1P_{263}6
    \rangle ^{2} \int d^{n}q \, \delta \left( q_a^{2} \right) \delta \left(
    q_b^{2} \right)
    \frac{{A^{scalar}_{1}}_{tree}{A^{scalar}_{2}}_{tree}}{\langle
    1a6 \rangle \langle 1b6\rangle} &
\end{align}
\noindent Again, thanks to the supersymmetric decomposition
$\left( \ref{supersymetricdecomposition} \right)$, we can identify
the supersymmetric cut amplitude and obtain the fermion cut amplitude.
Doing all the calculation, we get for these two amplitudes:
\begin{align}
    A_{6}^{fermion /\caln=1}(---+++) = & \ \dsp i\frac{\left( e\sqrt{2}\right) ^{6}}{16 \pi^{2}} \sum_{\sigma(1,2,3)}\sum_{\sigma(4,5,6)} & \nonumber\\
    & \left( 2 \, d_{i}^{fermion/\caln=1} F_{2A}(s_{14},s_{452},{s}_{25},{s}_{36})+ \frac{e_{i}^{fermion/\caln=1}}{2} F_{1}(s_{63},s_{61},s_{425}) \right.& \nonumber\\
    & + \dsp  \ \ \left. \frac{e_{i}^{fermion/\caln=1}}{2}^{*} F_{1}(s_{25},s_{24},s_{136}) + \frac{g_{i}^{fermion/\caln=1}}{6} {I_{3}^{n}}^{3m}({s}_{14},{s}_{25},{s}_{36}) \right)&
\end{align}
\noindent where:
\begin{align}
    d_{i}^{fermion} & = \dsp -  \frac{ 1}{[31]\langle 45 \rangle [ 1 p_{425} 4]^{2}}
    \frac{[1p_{425}2]^{2}[6p_{425}4]^{2} + s_{425}^{2} \langle 24 \rangle^{2} [16]^{2}}{[1p_{425}5][3p_{425}4]} \label{d_fermion} & \\
    e_{i}^{fermion} & = \dsp  2 \frac{\langle 2 p_{425} 3\rangle ^{2} [16]^{2} s_{425}}
    {\langle 4p_{425}1 \rangle \langle 5p_{425}3 \rangle \langle 5p_{425}1 \rangle \langle 4p_{425}3 \rangle } \frac{\langle 31 \rangle}{ [31] s_{31}}&\\
    {e_{i}^{fermion}}^{*} & = \dsp 2 \frac{[6p_{613}5]^{2} \langle 42 \rangle^{2} s_{613}}{[1p_{613}4][1p_{613}5][3p_{613}4][3p_{613}5]}\frac{[54]}{\langle 54 \rangle s_{54}} &\\
    g_{i}^{fermion} & =  -2g_{i}^{scalar}
    &
\end{align}
\noindent Again, we get a full agreement with~\cite{binoth:2007}
for the amplitude $A_{6}^{fermion}(---+++)$. We can observe a
factor "2", except for the coefficient of the two adjacent box
between the coefficients of the two amplitudes
$A_{6}^{fermion}(---+++)$ and $A_{6}^{scalar}(---+++)$. This
factor "2" comes from the fact that for a fermion loop, there are
two currents. These two currents give rise to a factor "2" except
in the case of the two adjacent mass box, where they give rise to
a sum two terms (eq. (\ref{d_fermion})).
\begin{align}
    d_{i}^{\caln=1} & = \dsp - \frac{1}{[31]\langle 45\rangle}\frac{[6p_{425}2]^{2}}{[1p_{425}5][3p_{425}4]} & \\
    e_{i}^{\caln=1} & = \dsp 2d_{i}^{\caln=1}  &\\
    {e_{i}^{\caln=1}}^{*} & = e_{i}^{\caln=1} &\\
    g_{i}^{\caln=1} & = 0  &
\end{align}
\noindent In the ${\caln=1}$ supersymmetric QED, the coefficients are simpler. Since there are only six
photons, the two kinds of one mass four point boxes are
related by parity. This leads to the equality between the coefficients
${e_{i}^{\caln=1}}^{*} = e_{i}^{\caln=1}$. With eight photons or more, this
becomes wrong. We can also point out that all Gram determinants
disappear.

\hfil

\subsection{ Absence of triangles in $A_6^{\caln=1}$}

\hfil

In the $\caln=1$ supersymmetric amplitude $A_6^{\caln=1}$, the
fact that there are no triangles is probably just an accident. But
we can make the following statement: for one loop, in
$QED^{\caln=1}$ theory, the $N$ photons NMHV helicity amplitudes
will have no triangle function.  In fact, as we have only three
photons with a negative helicity, each mass of the triangle is
formed with one photon with a negative helicity. The relation
$\left( \ref{arbre4QEDS} \right)$ shows the linearity between the
MHV tree amplitude in QED and in scalar QED. So we obtain directly
using \cite{Forde:2007mi} that:
\begin{align}
    g^{\caln=1} & = g^{fermions} +2 g^{scalar} &
\end{align}
\noindent But now if we have at least four photons with a negative
helicity, one mass of the triangle will be formed by two photons
with a negative helicity and the tree amplitude corresponding to
this mass will not be a MHV tree. The problem comes from the fact
that the relation linking NMHV tree amplitude in QED and scalar
QED is not linear but affine. This affine coefficient is the
contribution of the triangle to the supersymmetric $\caln=1$ amplitude
and there is no reason that the coefficients in front of the
triangles become zero. So we can conjecture that at one loop, in
$QED^{\caln=1}$ theory, the $N>6$ photons Next to Next to MHV (NNMHV) helicity amplitudes
will have triangles.

\hfil

\section{Numerical results}

\hfil

In this section, we present numerical results for the different
six-photon amplitudes. We have built a fortran code using the
GOLEM library for the different four point and three point
functions. To be consistent with previous results, we use the
kinematics defined by Nagy and Soper in ref. \cite{nagy}. First of
all, we recall this kinematics: the reaction $
\gamma^{\lambda_1}(k_1)+\gamma^{\lambda_2}(k_2) \rightarrow
\gamma^{\lambda_3}(k_3)+\gamma^{\lambda_4}(k_4)+\gamma^{\lambda_5}(k_5)+\gamma^{\lambda_6}(k_6)$
is considered, where $k_i$ is the four-momentum of the photon $i$
and $\lambda_i$ its helicity, the four-momenta fulfill $k_1 + k_2
= k_3 + k_4 + k_5 + k_6$. In the center of mass frame
$\vec{k}_1+\vec{k}_2 = \vec{0}$, $\vec{k}_1$ is along the -z-axis,
an arbitrary phase space point is chosen:
\begin{equation}
\left\{ \begin{array}{l}
    \overrightarrow{k_{3}} = (33.5,15.9,25.0) \\
    \overrightarrow{k_{4}} = (-12,5,15.3,0.3) \\
    \overrightarrow{k_{5}} = (-10.0,-18.0,-3.3) \\
    \overrightarrow{k_{6}} = (-11.0,-13.2,-22.0)
\end{array} \right.
\end{equation}
\noindent Then new final momentum configurations is generated by
rotating the final state through angle $\theta$ about the y-axis.
For all the plots of this section, we take $\alpha = e^{2}/4\pi =
1$. The helicities "+" and "-" refer always to ingoing photons.

\subsection{The MHV amplitudes.}

\hfil

\begin{figure}[httb!]
\centering
\includegraphics[scale=0.5]{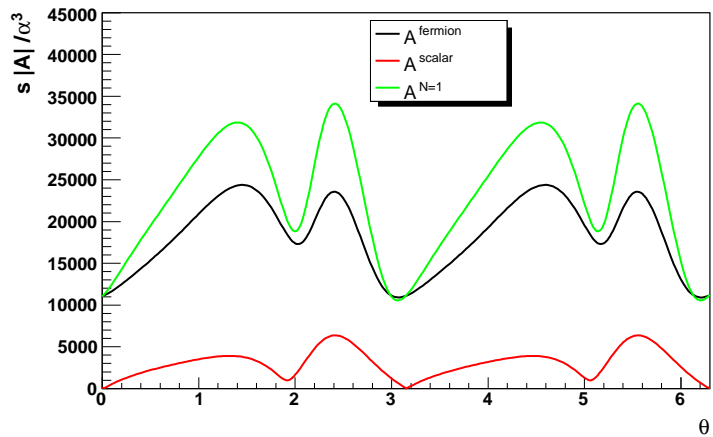}
\caption{\scriptsize \textit{The MHV six-photon amplitudes with
the Nagy and Soper \cite{nagy} configuration for the three
theories. Note that the curve for $A^{scalar}$ amplitude has a
minimum which is not zero.}} \label{plot_number1}
\end{figure}

In the figure \ref{plot_number1}, we plot the module of the
MHV amplitude for QED, scalar QED and $\caln=1$ supersymmetric QED
(respectively the formula (\ref{mhv_fermion}) , (\ref{mhv_scalar})
and (\ref{mhv_n=1}) ) against the variable $\theta$. Note that
these formula have been derived assuming that all the photons are
ingoing. In order to match previous results of the references
\cite{nagy}, \cite{binoth:2007}, we compute
$A_{6}^{fermion/scalar/\caln=1}(k_1,k_2,-k_3,-k_4,-k_5,-k_6)$ with the helicities $\lambda_1 = -$, $\lambda_2 = -$,
$\lambda_3 = +$, $\lambda_4 = +$, $\lambda_5 = +$, $\lambda_6 = +$.

\noindent All these MHV amplitudes are $\pi$ periodic.

\hfil

\subsection{The NMHV amplitudes}

\hfil

\begin{figure}[httb!]
\centering
\includegraphics[scale=0.5]{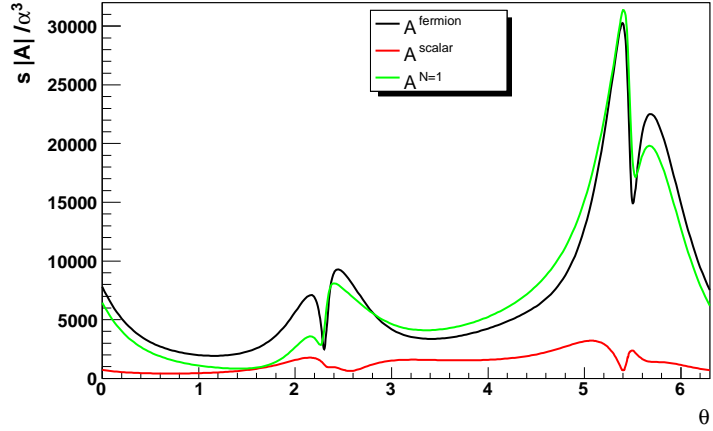}
\caption{\scriptsize \textit{NMHV six-photon amplitudes with the
Nagy and Soper \cite{nagy} configuration for the three QED.}}
\label{plot_number2}
\end{figure}

In figure \ref{plot_number2}, with the same configuration than in
the last section, we plot the NMHV six-photon amplitudes for the
three theories against $\theta$. We compute
$A_{6}^{fermion/scalar/\caln=1}(k_2,-k_3,-k_6,k_1,-k_4,-k_5)$ with
the helicities $\lambda_1 = +$, $\lambda_2 = -$, $\lambda_3 = -$,
$\lambda_4 = +$, $\lambda_5 = +$, $\lambda_6 = -$.

\noindent In this case, the amplitudes are not $\pi$ periodic. The
dips appearing in the curves, are related to the Landau
singularities called the "double parton scattering" \cite{nagy}.

\hfil

\subsection{Double parton scattering}

\hfil

The Landau equations give the necessary conditions for a Feynman
diagram to have a singularity. In the case of the six-photon
amplitudes, since all internal and external particles are massless,
three types of singularities can appear. Two are the well known
soft and collinear singularities, the other corresponds to the so
called double parton scattering. In this case, four propagators
are one mass shell, these propagators are adjacent by pair.

\hfil

So here we explain rapidly what is the "double parton
scattering" kinematics.

\hfil
\begin{figure}[httb!]
\centering
\includegraphics[width=6cm]{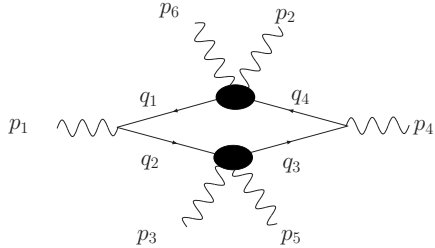}
\caption{\scriptsize \textit{Double parton scattering
configuration: $p_{1},p_{4}$ are ingoing photons and
$p_{2},p_{3},p_{5},p_{6}$ are outgoing photons.}}
\label{dessin_number6}
\end{figure}

The two ingoing photons 1 and 4 split into a fermion anti-fermion
collinear pairs, each fermion scatter with an anti-fermion to give
a photon pair with no transverse momentum in the center of mass frame $\vec{p}_1 +
\vec{p}_4 = \vec{0}$ (c.f. figure \ref{dessin_number6}). In the
configuration of "double parton scattering", the two propagators
$q_1$ and $q_2$ are collinear to the external leg $p_1$ and the
two other $q_3$ and $q_4$ are collinear to $p_4$:
\begin{equation}
\left\{ \begin{array}{l}
    q_{1} =  -x p_{1} \\
    q_{2} =  (1-x) p_{1} \\
    q_{3} =  -y p_{4} \\
    q_{4} =  (1-y) p_{4}
\end{array} \right.
\end{equation}
\noindent Solving the Landau equations, we find that the
conditions to have a double parton scattering singularity are:
\begin{equation}
\left\{ \begin{array}{l}
    det(S) \rightarrow 0 \\
    s_{35},s_{26} >0 \\
    s_{135},s_{435}<0
\end{array} \right.
\end{equation}
with
\begin{align}
    det(S) = s_{135}s_{435} -s_{35}s_{26}
\end{align}

In the center of mass frame $\vec{p}_1 + \vec{p}_4 = \vec{0}$,
$det(S)=s_{14} \, k_t^2$ where $k_t^2$ is the square of the
transverse momentum of the photon photon pairs 2,6 and 3,5.

\begin{figure}[httb!]
\centering
\includegraphics[scale=0.5]{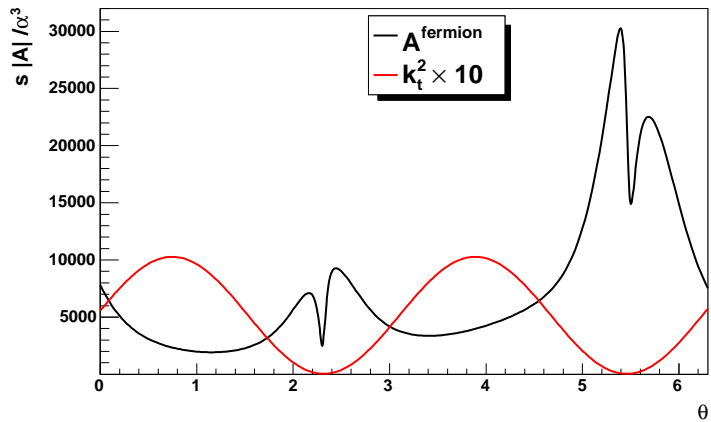}
\caption{\scriptsize \textit{Localisation of a Landau singularity.
Note that the red curve describing the $k_t^2$ of the photon pairs
2,6 and 3,5 does not reach zero.}} \label{plot_number3}
\end{figure}

\noindent In the fig. \ref{plot_number3}, we plot the NMHV QED
amplitudes as a function of $\theta$. On the top of that, we
surimpose $k_t^2$, normalized in such way that the curve is
visible. We note that the dips appear in the region where
$k_t^2$ is minimum. In this case, the minimum of $k_t^2$ is
different from zero because we are not sitting on the singularity.
A more detail study, will be presented in a forthcoming
publication \cite{a_venir}.

\section{Conclusion}

\hfil

In this paper, we have obtained all six-photon helicity amplitudes
in QED, scalar QED and $\caln=1$ supersymmetric QED. Those amplitudes
are linked among themselves by the relation $\left(
\ref{supersymetricdecomposition} \right)$. To calculate them, we
used the powerful unitarity-cut technics and we got very compact
expressions. More work is required to understand quantitatively
the behavior of these amplitudes and especially if the numerator
can regularize the double parton scattering singularity in a
kinematics where it shows up, this will be presented elsewhere
\cite{a_venir}.

\hfil

\section*{Acknowledgements}

We want to thank T. Binoth, G. Heinrich and C. Schubert for
collaboration during the early stages of this work and for useful
discussions about ref. \cite{binoth:2007}. We also would like to
thank E. Pilon and Ninh Le Duc for useful discussions on the
Landau singularity and P. Aurenche for a careful reading of the
manuscript.

\begin{appendix}
\renewcommand{\theequation}{\Alph{section}.\arabic{equation}}
\setcounter{equation}{0}

\section{Scalar integrals \label{scalar_int}}

In this appendix, for sake of completeness, the definition of
master integrals used in this paper is recalled, more details can
be found in \cite{Binoth:2001vm}. We also give $det(G)$ the
determinant of the Gram matrix $G_{ij} = 2 p_{i}.p_{j}$ built with
the external four momentum and $det(S)$ the determinant of the
kinematical S-matrix defined by $S_{ij} = \left( q_{j} - q_{i}
\right)^{2}$ where the $q_i$ are the four momentum flowing in the
propagators.

\subsection{Three mass three point function}

\begin{equation*}
\parbox{2.2cm}{\includegraphics[width=2.2cm]{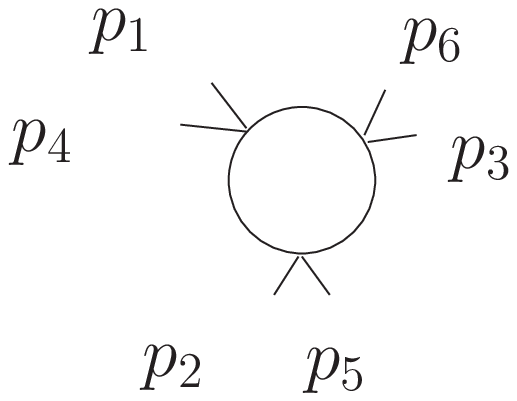}}
\quad \quad \quad \quad \quad \quad \quad \quad
\left\{\begin{array}{l}
    m_{1}^{2} = s_{14} \\
    m_{2}^{2} = s_{25} \\
    m_{3}^{2} = s_{36}
\end{array} \right.
\end{equation*}

\begin{align}
    I_{3}^{n} \left( m_{1}^{2},m_{2}^{2},m_{3}^{2} \right) = &
    \dsp \frac{1}{\sqrt{\Delta}} \left\{ \left ( 2 Li_{2} \left( 1-
    \frac{1}{y_{2}}\right) + 2 Li_{2} \left( 1-
    \frac{1}{x_{2}}\right) + \frac{\pi^{2}}{3} \right) \right.& \nonumber \\
    & \left.\dsp +\frac{1}{2}\left ( \ln^{2}\left( \frac{x_{1}}{y_{1}}
    \right)+ \ln^{2}\left( \frac{x_{2}}{y_{2}} \right) + \ln^{2}\left( \frac{x_{2}}{y_{1}}
    \right) - \ln^{2}\left( \frac{x_{1}}{y_{2}} \right) \right) \right\} &
    \label{3p3m}
\end{align}

\noindent where:

\begin{align}
    x_{1,2} & = \dsp \frac{m_{1}^{2} + m_{2}^{2} - m_{3}^{2} \pm \sqrt{\Delta} }{ 2
    m_{1}^{2}}& \\
    y_{1,2} & = \dsp \frac{m_{1}^{2} - m_{2}^{2} + m_{3}^{2} \pm \sqrt{\Delta} }{ 2
    m_{1}^{2}}& \\
    \Delta & = \dsp m_{1}^4 + m_{2}^4 + m_{3}^4 -2 m_{1}^{2} m_{2}^{2} -2 m_{1}^{2}m_{3}^{2} -2
    m_{2}^{2}m_{3}^{2} -i \ \textrm{sign}(m_{1}^{2}) \ \epsilon &
\end{align}

\noindent The formula (\ref{3p3m}) is valid in all kinematical
regions because of the small imaginary part $i \ \epsilon $:
\begin{eqnarray}
    \sqrt{\Delta \pm i \epsilon}  & = & \left\{ \begin{array}{l}
    \sqrt{\Delta} \pm i \epsilon \, , \;
    \Delta \geq 0 \\
    \pm i \, \sqrt{-\Delta} \, , \; \Delta \leq 0
    \end{array} \right.
\end{eqnarray}
\noindent The two determinants are given by the relations:
\begin{align}
    det(G_{3m}) & = m_{1}^{2}m_{2}^{2}-\left( m_{1}.m_{2}\right)^{2} = - \frac{\Delta}{4}&\\
    det(S_{3m}) & = 2m_{1}^{2}m_{2}^{2}m_{3}^{2} &
\end{align}

\subsection{Four point functions}

\subsubsection{With zero mass}

\begin{equation*}
\parbox{2.2cm}{\includegraphics[width=2.2cm]{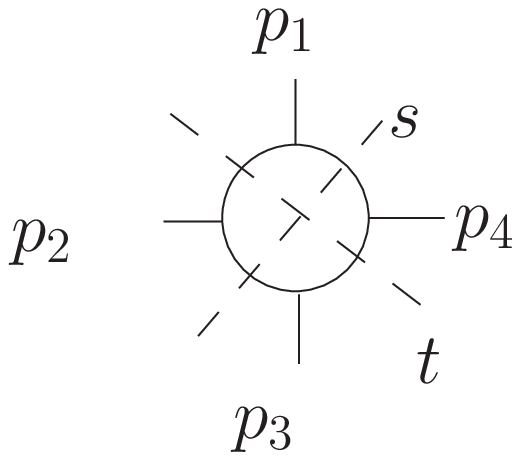}}
\quad \quad \quad \quad \quad \quad \quad \quad
\left\{\begin{array}{l}
    s = s_{12} \\
    t = s_{14} \\
    u = s_{13}
\end{array} \right.
\end{equation*}

\begin{align}
    I_{4}^{n} \left( s,t \right) = & \dsp \frac{2}{st} \frac{r_{\Gamma}}{\epsilon
    ^{2}}\left\{ (-s)^{-\epsilon} + (-t)^{-\epsilon} \right\} -\frac{2}{ st} F_{0} (s,t) &
\end{align}
\noindent where:
\begin{align}
    F_{0} (s,t) & = \dsp \frac{1}{2}\left\{  ln^{2} \left( \frac{s}{t} \right) + \pi^{2} \right\}&
\end{align}
\noindent The determinants are given by:
\begin{align}
    det(G_{0}) & = -2st(s+t) = 2stu &\\
    det(S_{0}) & = (st)^{2} = \langle 24342 \rangle ^{2} &
\end{align}

\hfil

\subsubsection{With one mass}

\begin{equation*}
\parbox{2.2cm}{\includegraphics[width=2.2cm]{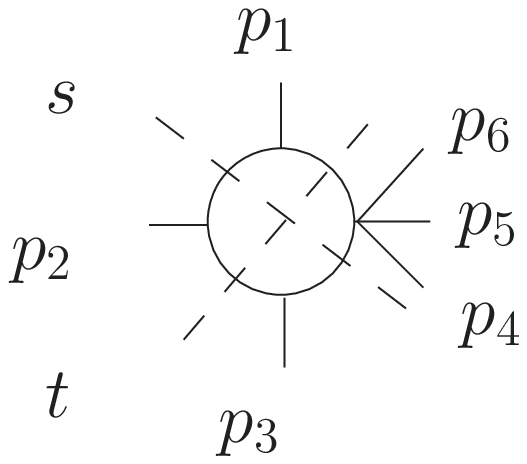}}
\quad \quad \quad \quad \quad \quad \quad \quad
\left\{\begin{array}{l}
    s = s_{12} \\
    t = s_{23} \\
    u = s_{13} \\
    m^{2} = s_{456}
\end{array} \right.
\end{equation*}

\begin{align}
    I_{4}^{n} \left( s,t,m^{2} \right) = & \dsp \ \  \frac{r_{\Gamma}}{st \epsilon
    ^{2}}\left\{ \left( (-s)^{-\epsilon} + (-t)^{-\epsilon} \right)
    + \left( (-s)^{-\epsilon} - (-m^{2})^{-\epsilon} \right)
    + \left( (-t)^{-\epsilon} - (-m^{2})^{-\epsilon} \right)
    \right\} & \nonumber \\
    & \dsp - \frac{2}{st} F_{1} \left(s,t,m^{2} \right) &
\end{align}
\noindent where:
\begin{align}
    F_{1} \left(s,t,m^{2} \right)& \dsp = Li_{2} \left(1 - \frac{m^{2}}{s}\right) + Li_{2} \left(1 -
    \frac{m^{2}}{t} \right) - Li_{2} \left(- \frac{s}{t} \right)
    - Li_{2} \left(- \frac{t}{s} \right) & \\
    & = F_{0} (s,t) +
    \left\{ Li_{2} \left(1 - \frac{m^{2}}{s}
    \right) + Li_{2} \left(1 - \frac{m^{2}}{t} \right) - \frac{\pi^{2}}{3}
    \right\} &
\end{align}
\noindent The determinants are given by:
\begin{align}
    det(G_{1}) & = \ -2st\left(s+t-m^{2}\right) = 2stu &\\
    det (S_{1}) & = \ (st)^{2} =  \langle 2m3m2 \rangle ^{2} &
\end{align}

\subsubsection{With two adjacent masses}

\begin{equation*}
\parbox{2.2cm}{\includegraphics[width=2.2cm]{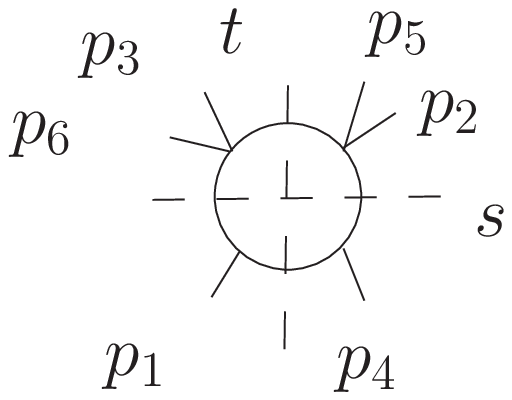}}
\quad \quad \quad \quad \quad \quad \quad \quad
\left\{\begin{array}{l}
    s = s_{14} \\
    t = s_{425} \\
    m_{1}^{2} = s_{25} \\
    m_{2}^{2} = s_{36}
\end{array} \right.
\end{equation*}

\begin{align}
    I_{4}^{n} \left( s,t,m_{1}^{2},m_{2}^{2} \right) = & \dsp \ \ \frac{r_{\Gamma}}{(st) \epsilon
    ^{2}}\left\{ (-s)^{-\epsilon}
    + \left( (-t)^{-\epsilon} - (-m_{1}^{2})^{-\epsilon} \right)
    + \left( (-t)^{-\epsilon} - (-m_{2}^{2})^{-\epsilon} \right) \right\} & \nonumber \\
    & \dsp -\frac{2}{st} F_{2A} \left(s,t,m_{1}^{2},m_{2}^{2} \right) &
\end{align}

\noindent where:

\begin{align}
    F_{2A} \left(s,t,m_{1}^{2},m_{2}^{2} \right) & = \dsp Li_{2} \left(1 - \frac{m_{1}^{2}}{t} \right) + Li_{2} \left(1 - \frac{m_{2}^{2}}{t}\right)
    + \frac{1}{2} ln \left( \frac{s}{t} \right) ln \left( \frac{m_{2}^{2}}{t} \right)+
    \frac{1}{2} ln \left( \frac{s}{m_{2}^{2}} \right) ln \left( \frac{m_{1}^{2}}{t} \right)&
\end{align}

\noindent The determinants are given by:
\begin{align}
     det(G_{2A}) & = \ -2s\left(m_{1}^{2}m_{2}^{2} -t(m_{1}^{2}+m_{2}^{2}-s-t)\right) = \ -2s\langle 1m_{1}4m_{2}1
     \rangle&\\
     det (S_{2A}) & = \ (st)^{2} &
\end{align}

\subsubsection{With two opposite masses}

\begin{equation*}
\parbox{2.2cm}{\includegraphics[width=2.2cm]{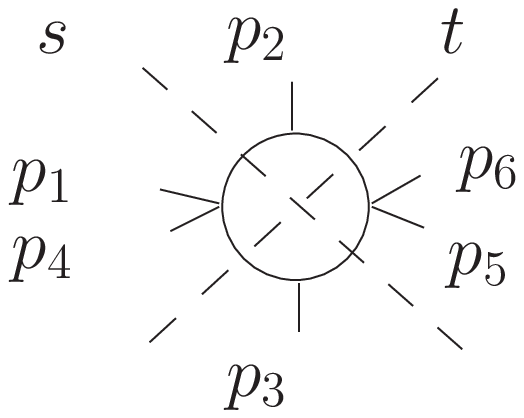}}
\quad \quad \quad \quad \quad \quad \quad \quad
\left\{\begin{array}{l}
    s = s_{143} \\
    t = s_{243} \\
    u = s_{23} \\
    m_{1}^{2} = s_{14} \\
    m_{2}^{2} = s_{56}
\end{array} \right.
\end{equation*}

\begin{align}
    I_{4}^{n} \left( s,t,m_{1}^{2},m_{2}^{2} \right) = & \dsp \frac{r_{\Gamma}}{(st-m_{1}^{2}m_{2}^{2}) \epsilon
    ^{2}}\left\{ \left( (-s)^{-\epsilon} - (-m_{1}^{2})^{-\epsilon} \right)
    + \left( (-s)^{-\epsilon} - (-m_{2}^{2})^{-\epsilon} \right)
    \right\} & \nonumber \\
    & + \dsp \frac{r_{\Gamma}}{(st-m_{1}^{2}m_{2}^{2}) \epsilon
    ^{2}}\left\{ \left( (-t)^{-\epsilon} - (-m_{1}^{2})^{-\epsilon} \right) + \left( (-t)^{-\epsilon} - (-m_{2}^{2})^{-\epsilon} \right)
    \right\} & \nonumber \\
    &  \dsp -\frac{2}{st - m_{1}^{2}m_{2}^{2}} F_{2B} \left(s,t,m_{1}^{2},m_{2}^{2} \right) &
\end{align}

\noindent where:

\begin{align}
    F_{2B} \left(s,t,m_{1}^{2},m_{2}^{2} \right) = & \dsp  \  - Li_{2} \left(1 - \frac{m_{1}^{2}m_{2}^{2}}{st}
    \right) + Li_{2} \left(1 - \frac{m_{1}^{2}}{s} \right) & \nonumber \\
    & \dsp + Li_{2} \left(1 - \frac{m_{2}^{2}}{s} \right)
    + Li_{2} \left(1 - \frac{m_{1}^{2}}{t} \right) + Li_{2} \left(1 - \frac{m_{2}^{2}}{t} \right) +  \frac{1}{2} ln^{2} \left( \frac{s}{t}\right) & \\
    = & \dsp \ F_{1} \left(s,t,m_{1}^{2} \right) + F_{1} \left(s,t,m_{2}^{2}
    \right)- F_{0} (s,t) - \left\{ Li_{2} \left( 1 - \frac{m_{1}^{2}m_{2}^{2}}{st} - \frac{\pi^{2}}{6}\right) \right\} &
\end{align}

\noindent The determinants are given by:
\begin{align}
    det(G_{2B}) & = -2\left(m_{1}^{2}m_{2}^{2}-st \right) \left(m_{1}^{2}+m_{2}^{2}-s-t\right) = 2u\left(st-m_{1}^{2}m_{2}^{2}\right) &\\
    det(S_{2B}) & = \left( st -m_{1}^{2}m_{2}^{2} \right)^{2} = \langle 2m_{1}3m_{1}2 \rangle ^{2} = \langle 2m_{2}3m_{2}2 \rangle ^{2} &
\end{align}

\hfil

\section{Proof of rule (\ref{rule2}) \label{p_rule2}}

\hfil

In this appendix, we want to prove that, in the case of the box
with two opposite masses, the helicities of the two opposite
massless legs must be the same otherwise the coefficient in front
is zero. To do that, we consider the following box integrals where
the helicities of the two opposite massless legs is different.
\begin{equation*}
\parbox{3cm}{\includegraphics[width=3cm]{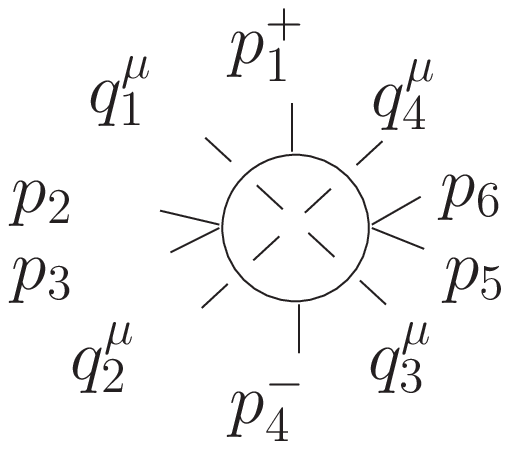}}
\quad \quad \quad \quad \quad \quad \quad \quad
\left\{\begin{array}{l}
    \forall i \in [1..6], \  p_{i}^{2} = 0 \\
    p_{23} = p_{2}+p_{3} \\
    p_{56} = p_{5}+p_{6}
\end{array} \right.
\end{equation*}
\noindent We assume that the helicity of the photon $p_{1}$ is
positive and the helicity of the photon $p_{4}$ is negative. Using
the four cuts technics, the coefficient, called $C$, in front of
this master integral is given by, in scalar QED:
\begin{align}
    C \propto \sum_{i=a,b} \varepsilon_{1}^{+}.{q_{1}}_{i}
    \varepsilon_{2}^{-}.{q_{2}}_{i} \label{coefficient}
\end{align}
where $ {q_{1}}_{i}$ and ${q_{2}}_{i}$ are the solution of the
four cuts conditions:
\begin{align}
    \delta(q_{1}^{2})  =0 \label{equation1} \\
    \delta(q_{2}^{2})  =0 \label{equation2} \\
    \delta(q_{3}^{2})  =0 \label{equation3} \\
    \delta(q_{4}^{2})  =0 \label{equation4}
\end{align}
\noindent So first we solve this system and after we will
calculate $\left( \ref{coefficient} \right)$.

\hfil

We choose as a base of the four-dimension Minkowski space: $ B =
\left\{ p_{1}^{\mu}, p_{4}^{\mu}, \langle 1 \gamma^{\mu} 4 \rangle
, \langle 4 \gamma^{\mu} 1 \rangle \right\}$. In our case,
$q_{1}^{\mu}$ can be taken as a four-dimension vector, therefore,
we can project it on the base $ B $:
\begin{align}
    q_{1}^{\mu} = a \ p_{1}^{\mu} + b \ p_{4}^{\mu} + \frac{c}{2} \langle 1 \gamma^{\mu} 4
    \rangle + \frac{d}{2} \langle 4 \gamma^{\mu} 1 \rangle
\end{align}
\noindent So to know the vector $q_{1}^{\mu}$, we have to
calculate, the four coefficient $a,b,c$ and $d$.
The conditions  (\ref{equation1}) and (\ref{equation4})
impose :
\begin{align}
    \left( q_{1} - p_{1} \right)
    ^{2} =0  \Leftrightarrow 2 \left( p_{1}.q_{1} \right) = 0  \Leftrightarrow b = 0
\end{align}
\noindent The conditions (\ref{equation2}) and (\ref{equation3}) impose:
\begin{align}
    \left( q_{1} + p_{23} + p_{4} \right)
    ^{2} =0  \Leftrightarrow 2  p_{4}.\left(q_{1} +p_{23} \right) = 0  \Leftrightarrow a s_{14} + 2(p_{23}.p_{4}) =
    0 \Leftrightarrow a = -\frac{2(p_{23}.p_{4})}{s_{14}}
\end{align}
\noindent The first condition $\left( \ref{equation1} \right)$,
knowing $b=0$, imposes:
\begin{align}
    q_{1} ^{2} =0  \Leftrightarrow c \ d = 0  \Leftrightarrow c = 0 \
    \textrm{or} \ d= 0
\end{align}
\noindent If we assume that $d=0$, therefore the conditions (\ref{equation1}) and
(\ref{equation2}) impose that:
\begin{align}
    \left( q_{1}
    + p_{23} \right) ^{2} = 0 \Leftrightarrow 2 (p_{23}.q_1) = -s_{23} \Leftrightarrow c = \frac{\langle 4 p_{23} 1\rangle}{s_{14}}
\end{align}
\noindent and finally we obtain $q_{1}^{\mu} = \dsp -\frac{\langle
4P_{23} \gamma^{\mu} 14 \rangle}{2 s_{14}}$. \\
Else if we assume $c=0$, therefore the conditions (\ref{equation1}) and
(\ref{equation2}) impose that:
\begin{align}
    \left( q_{1}
    + p_{23} \right) ^{2} = 0 \Leftrightarrow 2 (p_{23}.q_1) = -s_{23} \Leftrightarrow d = \frac{\langle 1 p_{23} 4\rangle}{s_{14}}
\end{align}
\noindent and in this case we obtain $q_{1}^{\mu} = \dsp -\frac{[
4P_{23} \gamma^{\mu} 14 ]}{2 s_{14}}$. Finally according to the
four cuts technics, the loop momenta is found to be:
\begin{align}
\left\{\begin{array}{l}
    {q_{1}^{\mu}}_{a} = \dsp -\frac{\langle 4P_{23} \gamma^{\mu} 14 \rangle}{2 s_{14}} \\
    {q_{1}^{\mu}}_{b} = \dsp -\frac{[ 4P_{23} \gamma^{\mu} 14 ]}{2 s_{14}}
\end{array} \right. \label{q1}
\end{align}
\noindent From the formula (\ref{q1}), we can compute
${q_{2}}_{a/b}$ and we obtain:
\begin{align}
\left\{\begin{array}{l}
    {q_{2}^{\mu}}_{a} = \dsp {q_{1}^{\mu}}_{a} + p_{23}^{\mu} = \frac{\langle 4 \gamma^{\mu} P_{23}  14 \rangle}{2 s_{14}} \\
    {q_{2}^{\mu}}_{b} = \dsp {q_{1}^{\mu}}_{b} + p_{23}^{\mu} = \frac{[ 4 \gamma^{\mu} P_{23}14 ]}{2 s_{14}}
\end{array} \right. \label{q2}
\end{align}
\noindent We are now ready to compute the left hand side of
equation $\left( \ref{coefficient} \right)$ inserting the formula
$\left( \ref{q1} \right)$ and $\left( \ref{q2} \right)$, we obtain
directly that:
\begin{align}
    C \propto \varepsilon_{1}^{+}.{q_{1}}_{a}
    \varepsilon_{2}^{-}.{q_{2}}_{a}+ \varepsilon_{1}^{+}.{q_{1}}_{b}
    \varepsilon_{2}^{-}.{q_{2}}_{b} = 0
\end{align}
\noindent Therefore the hypothesis that the two photons $p_{1}$
and $p_{4}$ have two different helicities implies that the
coefficient in front of the two opposite mass integrals is zero.

\end{appendix}

\end{document}